\documentclass[10pt,a4paper,showpacs,pra,preprint,twocolumn,lengthcheck]{revtex4}%
\usepackage{amsfonts}
\usepackage{amsmath}
\usepackage{amssymb}
\usepackage{graphicx}%
\usepackage{natbib}
\usepackage{color}
\providecommand{\U}[1]{\protect\rule{.1in}{.1in}}

\begin{document}

\preprint{ }
\title{Non-adiabatic Fast Control of Mixed States based on Lewis-Riesenfeld 
Invariant}
\author{Mohammad-Ali Fasihi$^{1,2}$}
\altaffiliation{On leave of absence from Department of Physics, Azarbaijan University of Tarbiat Moallem, 53714-161, Tabriz, Iran.}
\author{Yidun Wan$^1$}
\author{Mikio Nakahara$^{1,3}$}
\affiliation{$^1$Research Center for Quantum Computing,\\ 
Interdisciplinary Graduate School of Science and Engineering, 
Kinki University, \\
3-4-1 Kowakae, Higashi-Osaka, Osaka 577-8502, Japan\\
$^2$Department of Physics, Azarbaijan University of Tarbiat Moallem,\\ 53714-161, Tabriz, Iran.\\
$^3$Department of Physics, Kinki University, \\
3-4-1 Kowakae, Higashi-Osaka, Osaka 577-8502, Japan\\}
\keywords{Lewis-Riesenfeld Invariant, Nonadiabatic Evolution, Quantum Computation}
\pacs{37.10.De, 32.80.Qk, 42.50.-p, 03.65.Ca, 03.65.Vf, 03.65.Yz, 03.67.-a, 03.65.Ta}

\begin{abstract}
We apply the inversely-engineered control method based on Lewis-Riesenfeld 
invariants to control mixed states of a two-level quantum system. We show 
that the inversely-engineered control passages of mixed states - and pure 
states as special cases - can be made significantly faster than the 
conventional adiabatic control passages, which renders the method applicable 
to quantum computation. We devise a new type of inversely-engineered control 
passages, to be coined the antedated control passages, which further speed up 
the control significantly. We also demonstrate that by carefully tuning the 
control parameters, the inversely-engineered control passages can be optimized
in terms of speed and energy cost. 

\end{abstract}
\maketitle
\date[dated:]{Sept. 30, 2011}




\section{Introduction}
Adiabatic control has been one of the main methods of quantum control, 
quantum information processing, and quantum computation, because it is able 
to evolve a quantum system against parameter fluctuations. In particular, 
Farhi {\it et al.} proposed the adiabatic quantum computation \cite{AQCsci2001}, a new model of quantum computation based on adiabatic control, which was then 
extended to quantum open systems \cite{SarandyAQCOSprl2005,Sarandy2007}. Note, however, 
that adiabatic control gains its robustness on the cost of losing the speed of control, which seriously limits its applicability to cases where fast control - in particular faster than the decoherence of the system - is of great concern, such as quantum information processing and quantum computing \cite{Rowe2002,Chen2010prl}. 

In an adiabatic control of a quantum system, the system remains in the instantaneous ground state of its time-dependent Hamiltonian during the entire evolution. The control parameters in the Hamiltonian are carefully designed such that the adiabaticity condition always holds, which usually results in very long execution time. While other ways of speeding up an adiabatic control are available in the adiabatic regime \cite{SCRAP2008prl,Berry2009,Muga2009,Chen2010prlb}, Chen {\it et al.} put forward a new method, the inversely-engineered control (IEC) based on Lewis-Riesenfeld invariants \cite{ChenPRA2010,ChenPRA2011a,ChenPRA2011b,ChenPRA2011c,ChenPRA2011d,Chen2011a}. Lewis-Riesenfeld invariants (LRI) were discovered by Lewis and Riesenfeld in 1969 to solve the time-dependent Schr\"odinger equations \cite{LR1969}. In IEC, the Hamiltonian, which controls the system 
evolution, is determined by a corresponding LRI, which is constructed according to the desired initial and final states of the system. The IEC of a system is not adiabatic in general, although the associated Hamiltonian of the system may happen to satisfy the adiabaticity condition. Chen and his collaborators have shown that it was plausible 
that IEC offered shortcuts to adiabatic control of the pure states of 
a two-level quantum system without heating the system \cite{ChenPRA2011a,ChenPRA2011b}.

In this paper, we study IEC in more details by applying it to mixed states of a two-level quantum system. We consider mixed states because they are more general than pure states and play the key role in NMR (Nuclear Magnetic Resonance) quantum computation. In fact, how to control mixed states efficiently and rapidly is a major problem in NMR computation. We show that the IEC does give rise to fast, {\it non}adiabatic control passages of a two-level system, compared with the viable adiabatic control passages of the system. Moreover, the lower bound of the control time of IEC is only constrained physically but not mathematically. In particular, IEC bears an optimization regarding the time and energy cost of the control.

The paper is organized as follows. We briefly review the LRI and IEC in the next section. Section III is the main part of this paper, where the IEC of a mixed state is investigated. Section IV is devoted to conclusions 
and outlook. Throughout the paper, $\hbar$ is set to one.
\section{Lewis-Riesenfeld invariant and invariant-based inverse Engineering}
A Lewis-Riesenfeld Invariant $I(t)$, defined as a dynamical invariant of a quantum system with a time-dependent Hamiltonian $H(t)$, is a Hermitian operator
defined
on the Hilbert space of the system, which satisfies the following
defining equation \cite{LR1969}, 
\begin{equation}
\mathrm{i}\frac{\partial I(t)}{\partial t}-[H(t),I(t)]=0.\label{eqLRI}
\end{equation}
The dynamical invariance of $I(t)$ implies $\mathrm{d}\langle I(t)\rangle/\mathrm{d}t=0$, which endows $I(t)$ with a spectral decomposition
\begin{equation*}
I(t)=\sum_n\lambda_n|\phi_n(t)\rangle\langle\phi_n(t)|,
\end{equation*}
where $\lambda_n$ is real and time-independent. A LRI of a system enables the direct integration of the time-dependent Schr\"odinger equation of the system, giving rise to the general solution in terms of the instantaneous eigenstates of $I(t)$ \cite{LR1969}, i.e.,
\begin{equation}
|\Psi(t)\rangle=\sum_n{ c_n(0) e^{i\alpha_n(t)}|\phi_n(t)}\rangle,
\end{equation}
where $\alpha_{n}$ is the Lewis-Riesenfeld phase
\begin{equation}\label{eqLRphase}
\alpha_{n}=\int_0^t\langle\phi_n(t')|
\left(\mathrm{i}\frac{\partial}{\partial t'}-H(t')\right)|\phi_n(t')\rangle\mathrm{d}t'.
\end{equation}

In an adiabatic control of a system with a Hamiltonian $H_0$, one usually 
applies external fields on the system, such that the total Hamiltonian becomes $H(t)$, and drives the system from an initial Hamiltonian $H(0)=H_0$ to the final one, $H(t_f)$, by keeping the system in the instantaneous ground state of $H(t)$. The corresponding final state is either an eigenstate or a superposition of the eigenstates of $H(t_f)$, depending on the purpose of the control. We know that associated to this Hamiltonian is a LRI $I(t)$, which satisfies Eq.~(\ref{eqLRI}). The converse is also true. That is, there exists a $H(t)$ that corresponds to a given $I(t)$, which indicates a way around the adiabaticity constraint. That is, rather than evolving a system adiabatically, one first constructs a LRI $I(t)$, which then yields a Hamiltonian $H(t)$ by Eq.~(\ref{eqLRI}), that drives the system in an instantaneous eigenstate $|\phi_n(t)\rangle$ of $I(t)$. States $|\phi_n(0)\rangle$ and $|\phi_n(t_f)\rangle$ must coincide with the designated initial and final states of the original adiabatic control. At any intermediate time $t$, however, $|\phi_n(t)\rangle$ is not an eigenstate of $H(t)$ in general. 
This summarizes the idea of IEC.

Here is a strategy to implement this idea on any system.
1) Write down the total Hamiltonian $H(t)$ of the system and the control field, which reduces to $H_0$ at $t=0$ and $t=t_f$, but keep all the time-dependent parameters undetermined. 2) Use the most convenient LRI $I(t)$ of $H(t)$, leaving the parameters undetermined either. 3) Plug both $H(t)$ and $I(t)$ into Eq.~(\ref{eqLRI}), which should lead to a group of ordinary differential equations of the parameters in $I(t)$ and those in $H(t)$. 4) Plug in the specified
initial and final conditions on the parameters in $I(t)$. 5) Find physically and mathematically reasonable interpolations of the parameters in $I(t)$, such that they comply with the initial and final conditions. 6) Finally, determine the parameters in $H(t)$ by solving the group of differential equations. The Hamiltonian $H(t)$ determined this way is the desired one. This is indeed the strategy taken in \cite{ChenPRA2011a,ChenPRA2011b}. Sarandy {\it et al.} adopts another strategy based on Lie algebras \cite{Sarandy2011}.
\section{Inversely-Engineered Control of the Mixed States of Two-Level Atom}\label{secIEC}
From now on, we focus on a two-level quantum system to study IEC in details. 
Explicitly, we have a spin-1/2 nucleus under a strong magnetic field and an
oscillating transverse radio frequency (RF) fields in our mind. This system is realized
in NMR for example.
In Sections \ref{secIEC}.1 to 3.3, we shall explicitly follow the steps of our strategy described in the end of the last section. 
\subsection{The Lewis-Riesenfeld Invariant}
The typical two-level NMR Hamiltonian under an external RF field reads
\begin{equation}\label{eqHgen}
H(t)=\frac{1}{2}\left( %
\begin{matrix}
-\omega_0+\omega & \begin{array}{c}
\Omega_R\left[\mathrm{e}^{-\mathrm{i}[(\omega-\omega_{\mathrm{RF}})t+\varphi]}\right.\vspace{1mm} \\ 
\left. +\mathrm{e}^{-\mathrm{i}[(\omega+\omega_{\mathrm{RF}})t-\varphi]}\right]
\end{array}  \\ 
& &\\ 
\begin{array}{c}
\Omega_R\left[\mathrm{e}^{\mathrm{i}[(\omega-\omega_{\mathrm{RF}})t+\varphi]}\right. \vspace{1mm} \\ 
\left. +\mathrm{e}^{\mathrm{i}[(\omega+\omega_{\mathrm{RF}})t-\varphi]}\right]
\end{array}  & \omega_0-\omega
\end{matrix} \right),
\end{equation}
where $\Omega_R$, $\omega_{\mathrm{RF}}$, $\omega_0$, $\omega$, and $\varphi$ are respectively the Rabi frequency, the frequency of the RF field, the Zeeman 
energy of the system, the frequency of the rotating frame, and a phase. It is customary to configure the system such that $\omega+\omega_{\mathrm{RF}}\gg\omega-\omega_{\mathrm{RF}}$, which leads to the rotating-wave approximation, in
which the highly oscillating terms due to $\omega+\omega_{\mathrm{RF}}$ in Eq.~(\ref{eqHgen}) are averaged to vanish. Moreover, one can put $\omega=\omega_{\mathrm{RF}}$. 

For the purpose of control, however, we can modulate the Rabi frequency, the $\omega$, and the phase $\varphi$ with time. Therefore, by defining the detuning frequency $\Delta(t)=\omega(t)-\omega_0$, the Hamiltonian of our concern hereafter is
\begin{equation}\label{eqH}
H =\frac{1}{2}
\left(
  \begin{array}{cc}
    \Delta(t) & \Omega_R(t) e^{i\varphi(t)} \\
    \Omega_R(t) e^{-\mathrm{i}\varphi(t)} & -\Delta(t) \\
  \end{array}
\right).
\end{equation}
This Hamiltonian has eigenvalues $E_\pm=\pm{h \Omega}/{2}$, where $\Omega=\sqrt{\Delta^2+\Omega_R^2}$ is the generalized Rabi frequency. At $t=0$ and $t=t_f$ when the RF field is turned off, $H(0)=-H(t_f)=H_0=\mathrm{diag}(\Delta(0),-\Delta(0))$. In the sequel, we assume $\varphi(t)\equiv 0$, which further simplifies the $H(t)$ and renders the adiabaticity condition of the system
\begin{equation}\label{condAdia}
\left|\frac{\Omega_R\dot{\Delta}-\dot{\Omega}_R\Delta}{\Omega^3}\right|\ll 1,
\end{equation}
and the instantaneous eigenstates of the Hamiltonian
\begin{equation}\label{eqHeigenvectors}
\begin{aligned}
& |E_+(t)\rangle=\cos{\left(\frac{\theta}{2}\right)}|0\rangle
+\sin{\left(\frac{\theta}{2}\right)}|1\rangle,\\
& |E_-(t)\rangle=\sin{\left(\frac{\theta}{2}\right)}|0\rangle-
\cos{\left(\frac{\theta}{2}\right)}|1\rangle,
\end{aligned}
\end{equation}
where $\theta=\arccos(\Delta/\Omega)$ is the mixing angle. $|0\rangle=(1,\ 0)^T$ and $|1\rangle=(0,\ 1)^T$ are the eigenstates of the Pauli matrix $\sigma_z$. The Hamiltonian
$H(t)$ in Eq.~(\ref{eqH}) may serve to control the system adiabatically if it meets the adiabaticity condition in Eq.~(\ref{condAdia}). Note that none of the
time-dependent parameters in $H(t)$ is fixed.
This concludes the step 1) of our strategy.

A reasonable LRI $I(t)$ associated with $H(t)$ in Eq.~(\ref{eqH}) must be employed to complete the step 2) of our strategy. We adopt the general form 
employed in \cite{ChenPRA2011b}, i.e.,
\begin{equation}\label{eqIt}
I(t)=\frac{\Omega_0}{2}\left(
                           \begin{array}{cc}
                             \cos \gamma & e^{i\beta} \sin \gamma \\
                           e^{-\mathrm{i}\beta}  \sin \gamma  & -\cos \gamma \\
                           \end{array}
                         \right),
\end{equation}
where $\gamma$ and $\beta$ are time-dependent parameters that in turn determines $\Omega_R(t)$ and $\Delta(t)$ in the Hamiltonian, while $\Omega_0$ is an auxiliary constant. 

To follow step 3), we substitute Eqs. (\ref{eqH}) and (\ref{eqIt}) into Eq.~(\ref{eqLRI}), which immediately leads to two coupled differential equations
\begin{equation}\label{eqOmegaDelta}
\begin{aligned}
\Omega_R &=\frac{\dot{\gamma}}{\sin\beta},\\
\Delta &= \Omega_R\cot\gamma\cos\beta-\dot{\beta}.
\end{aligned}
\end{equation}
In IEC, we need first to find $\gamma$ and $\beta$ without solving Eq.~(\ref{eqOmegaDelta}) and then insert them into Eq.~(\ref{eqOmegaDelta}) to obtain $\Omega_R$ and $\Delta$. We now proceed to do this to control the mixed states of our
two-level system.
\subsection{Inversely-Engineered Control of Mixed States}

To complete the step 4), we first find the instantaneous eigenstates of $I(t)$, as the two-level system evolves in a mixed state in the basis of the instantaneous eigenstates of $I(t)$. From Eq.~(\ref{eqIt}), the instantaneous eigenstates of $I(t)$ are obtained as
\begin{equation}\label{eqLeigevevtors}
\begin{aligned}
& |\phi_+(t)\rangle=\cos{\left(\frac{\gamma}{2}\right)}e^{i\beta}|0\rangle
+\sin{\left(\frac{\gamma}{2}\right)}|1\rangle,\\
& |\phi_-(t)\rangle=\sin{\left(\frac{\gamma}{2}\right)}|0\rangle-
\cos{\left(\frac{\gamma}{2}\right)}e^{-\mathrm{i}\beta}|1\rangle.
\end{aligned}
\end{equation}
One can verify that in this case, the Lewis-Riesenfeld phase in Eq.~(\ref{eqLRphase}) becomes 
\begin{equation}\label{eqLRphase2level}
\alpha_{\pm}(t)=\mp\frac{1}{2}\int_0^t\tilde{\Omega}(t')\mathrm{d}t',
\end{equation}
where $\tilde{\Omega}(t)=(\Delta+\dot{\beta})\cos\gamma+\dot{\beta}
+\Omega_R \sin \gamma \cos\beta$. Hence, the instantaneous mixed state is
\begin{align}
&\rho^{I}(t) =\sum_{i=\pm}{p_i|\phi_{i}(t)\rangle\langle \phi_{i}(t)|}\label{eqRhoI}\\
&= \frac{1}{2}\left(
\begin{array}{cc}
  1+(p_+-p_-) \cos\gamma & (p_+-p_-) \mathrm{e}^{\mathrm{i}\beta} \sin\gamma \\
  (p_+-p_-) \mathrm{e}^{-\mathrm{i}\beta} \sin\gamma & 1-(p_+-p_-) \cos\gamma
\end{array}\nonumber
\right)
\end{align}
where $\sum_{i=\pm}{p_i}=1$. Note that the density matrix is diagonal in the basis of the eigenstates $|\phi_{\pm}\rangle$ of $I(t)$ but not in the basis
$|E_{\pm}(t) \rangle$ in general. Since we would like to depict the trajectories of the evolution of the mixed states in the Bloch ball for illustrative purposes, we cast the density matrix $\rho^I(t)$ in the basis of Pauli matrices
as
\begin{align}
\rho^I(t)=& \frac{1}{2}I_2+\frac{p_+-p_-}{2}[\sin\gamma(t)\cos\beta(t)\sigma_x\nonumber\\
&-\sin\gamma(t)\sin\beta(t)\sigma_y+\cos\gamma(t)\sigma_z],\label{eqRhoIbloch}
\end{align}
where $I_2$ is the $2\times 2$ unit matrix. The corresponding Bloch vector is $(p_+-p_-)\times (\sin\gamma(t)\cos\beta(t),\ -\sin\gamma(t)\sin\beta(t),\ \cos\gamma(t))$. Clearly, if either $p_+$ or $p_-$ is zero, $\rho^I(t)$ reduces to a pure state and the Bloch vector sits on the Bloch sphere.

We shall consider the control passages of $\rho^I$ that cause a complete population inversion in the two levels of the atom. Seen in Eq.~(\ref{eqLeigevevtors}), the population is completely controlled by $\gamma$ because $\beta$ is merely a phase. Therefore, in order to achieve the population inversion, $\gamma$ must satisfy certain initial and final conditions, which, following from Eq.~(\ref{eqLeigevevtors}), should be $\gamma(0)=n\pi$ and $\gamma(t_f)=m\pi$, where $n$ and $m$ are respectively even and odd integers, or vice versa. Without loss of generality, we assume $\gamma(0)=\pi$ and $\gamma(t_f)=0$, such that $\rho^I(0)=\mathrm{diag}((1-p_++p_-)/2,(1+p_+-p_-)/2))$ and $\rho^I(t_f)=\mathrm{diag}((1+p_+-p_-)/2,(1-p_++p_-)/2))$, which are indeed the corresponding mixed states in terms of the two levels of $H_0$, manifesting the population inversion. Although the initial and final density matrices are diagonal, the intermediate ones are generally not. We can constrain $\gamma$ further because the RF field and hence the Rabi frequency $\Omega_R$ is turned on at $t=0$ and off at $t=t_f$, which sets $\dot{\gamma}(0)=\dot{\gamma}(t_f)=0$ by Eq.~(\ref{eqOmegaDelta}). Summarized in below are the necessary conditions for $\gamma(t)$
\begin{equation}\label{condGamma3rd}
\begin{aligned}
&\gamma(0)=\pi,\ \ &\gamma(t_f)=0,\\
&\dot{\gamma}(0)=0,\ \ &\dot{\gamma}(t_f)=0.
\end{aligned}
\end{equation}

We now examine $\beta$. As previously remarked, $\beta$ plays no role in the evolution of the population number. However, it enters the Bloch vector corresponding to $\rho^I$ and thus affects the evolution trajectories of the states. Furthermore, Eq.~(\ref{eqOmegaDelta}) shows that $\beta$ also influences the profiles of $\Omega_R$ and $\Delta$. That is, inappropriate $\beta$ may make $\Omega_R$ and $\Delta$ unphysical --- ruins the rotating wave approximation,
for example --- and may cause the control impractical. We fix two most obvious conditions for $\beta$ for the moment but leave other conditions for the subsequent sections because they are case-dependent. Since $\gamma(0)=\pi$ and $\gamma(t_f)=0$, $\Delta$ can diverge at $t=0$ and $t=t_f$ because of the $\cot\gamma$ in Eq.~(\ref{eqOmegaDelta}) unless this divergence is compensated by vanishing $\cos\beta$. This calls for the following two conditions.
\begin{equation}\label{condBetaBasic}
\beta(0)=\beta(t_f)=-\pi/2.
\end{equation} 
They are negative in order that $\lim\limits_{t\rightarrow 0,t_f}\Omega_R=0^+$, 
which is consistent with the positivity of $\Omega_R$ due to the assumption that $\varphi\equiv 0$.

It is time to proceed to step 5) to find parameters $\gamma(t)$ and $\beta(t)$ that satisfy the above conditions and certain extra conditions to be set. To this end, Ansatz about the functional forms of $\gamma$ and $\beta$ should be made. The simplest one is that both $\gamma$ and $\beta$ are polynomials of finite order in $t$. We shall take this ansatz for our study in all following sections but one, where it will be explicitly stated. It turns out that the power of IEC in speeding up the control of the system varies in different cases. Throughout the rest of the paper, we take $p_+=0.2$ and $p_-=0.8$ for definiteness. Our
analysis, however, applies to any states including pure states.
 \begin{figure}[h]
 \centering
 \includegraphics[scale=0.51]{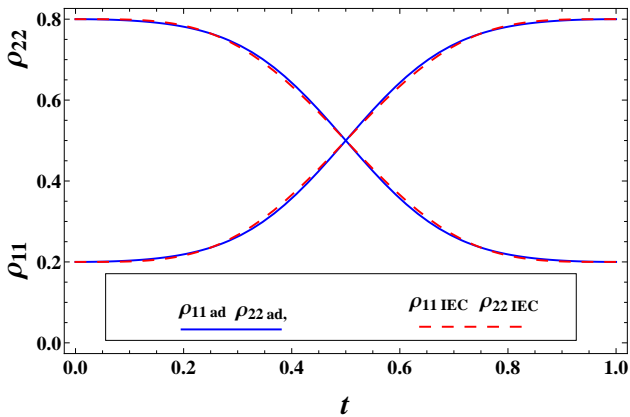}
  \caption{(Color online.) IEC (dashed lines) and adiabatic control (solid) passages of the mixed state in Eq.~(\ref{eqRhoI}). $t_f=1$.
   }
  \label{fig:mixedresult3}
 \end{figure}
\subsection{Third Order $\gamma$}
If we consider the conditions on $\gamma$ in Eq.~(\ref{condGamma3rd}) exclusively, then a third-order polynomial $\gamma(t)=\sum_{i=0}^3 a_i t^i$ is a good candidate for the functional form of $\gamma$ that interpolates these conditions. Solving for the coefficients $a_i$'s by substituting $\gamma$ into Eq.~(\ref{condGamma3rd}) shows that $\gamma(t)$ monotonically decreases from $\pi$ to $0$ with $t\in[0,t_f]$. This negativity of $\dot{\gamma}(t)$ indicates that $\sin\beta$ must be negative for $\Omega_R$ to remain positive by Eq.~(\ref{eqOmegaDelta}). As such, we assume $\beta(t)$ is also third-order in $t$ and accordingly add to Eq.~(\ref{condBetaBasic}) two more conditions on $\beta$, i.e., $\dot{\beta}(0)=3\pi/(2t_f)$ and $\dot{\beta}(t_f)=-3\pi/(2t_f)$\footnote{This condition is adopted from \cite{ChenPRA2011b} to compare the corresponding results for pure states therein.}, which simply fix the initial and final values of the detuning, such that $H(t_f)=-H(0)$. Plugging $\gamma$ and $\beta$ so obtained into Eq.~(\ref{eqRhoI}) gives the corresponding $\rho^I(t)$ with $t\in[0,t_f]$. 
The solid curves in Fig. \ref{fig:mixedresult3} show the control passages
$\rho^I_{11}(t)$ and $\rho^I_{22}(t)$. Step 5) is now completed.

We remark that in Fig.~\ref{fig:mixedresult3} and in all subsequent figures, the final time $t_f$ can have any positive value, such as microseconds or nanoseconds. In other words, mathematically, $t_f$ can be arbitrarily close to zero. Physically, of course, exceedingly small $t_f$ may set the energy/frequency scale of $\Omega_R$ and $\Delta$ too high to be practical or realizable. Another physical lower bound of $t_f$ roots in the uncertainty principle \cite{bh}.
\begin{figure}[h]
\begin{center}
\includegraphics[scale=0.6]{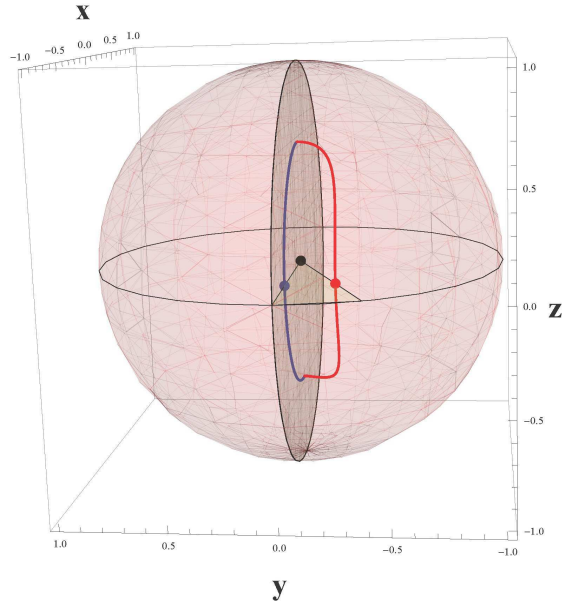}
\end{center}
 \caption{(Color online.) Bloch vector trajectories of the mixed state under the IEC (outside the $xz$-plane) and the adiabatic control (lying in the $xz$-plane), corresponding to Fig.\ref{fig:mixedresult3}. Three dots represent respectively the sphere center and where the IEC and adiabatic trajectories intersect the $xy$-plane. The solid straight lines connecting the three dots help with the visualization of the spatial relations of the two trajectories. The shaded disk in the sphere is the $xz$-plane.
 }
 \label{fig:mixed3bloch}
\end{figure}

The parameters $\Omega_R$ and $\Delta$ calculated from $\gamma$ and $\beta$, corresponding to the control passages in Fig.~\ref{fig:mixedresult3}, happen to satisfy the adiabaticity condition Eq.~(\ref{condAdia}). Hence, one can directly perform an adiabatic control of the mixed state without recourse to LRI. The adiabatically evolving density matrix reads
\begin{align}
\rho^{ad}(t) &=\sum\limits_{i=\pm}p_i|E_i(t)\rangle\langle E_i(t)|\label{eqRhoAdia}\\
& =\frac{1}{2}I_2+\frac{p_+-p_-}{2}[\sin\theta\sigma_x+\cos\theta\sigma_z],\nonumber
\end{align}
where the second line is the Bloch vector decomposition, and $\theta=\theta(t)$ is the mixing angle. The matrix elements $\rho^{ad}_{11}(t)$ and $\rho^{ad}_{22}(t)$
of the adiabatic control passages 
are depicted as solid lines in Fig.~\ref{fig:mixedresult3}. Note that they almost coincide with the IEC passages. In this case, the IEC is as slow as the adiabatic control. Nonetheless, the trajectories of the mixed state evolving under the IEC and adiabatic control respectively are rather different in the Bloch ball as visualized in Fig.~\ref{fig:mixed3bloch}. It is clear from Eqs. (\ref{eqRhoIbloch}) and (\ref{eqRhoAdia}) that the adiabatic Bloch vector is confined within the $xz$-plane but the IEC one is not.

In theory, the Hamiltonian inversely-engineered from a LRI may violate the adiabaticity condition Eq.~(\ref{condAdia}). In such a scenario, if one still insisted on performing an adiabatic control with this particular Hamiltonian, one would have to accomplish it in literally an infinite amount of time. Consequently, when the inversely-engineered Hamiltonian disallows an adiabatic control in an acceptable time duration $t_f$, IEC outmatches adiabatic control significantly because its control time is bounded below only by the physical realizability of the control parameters and the uncertainty principle. This motivates us to investigate how to accelerate IEC more and optimize it in the nonadiabatic regime. Recall that $\beta$ does not directly affect the evolution of the population. Thus,
we should first manipulate $\gamma$ and then $\beta$ if necessary.
\subsection{Fourth Order $\gamma$}
\begin{figure}[h]
\begin{center}
\includegraphics[scale=0.46]{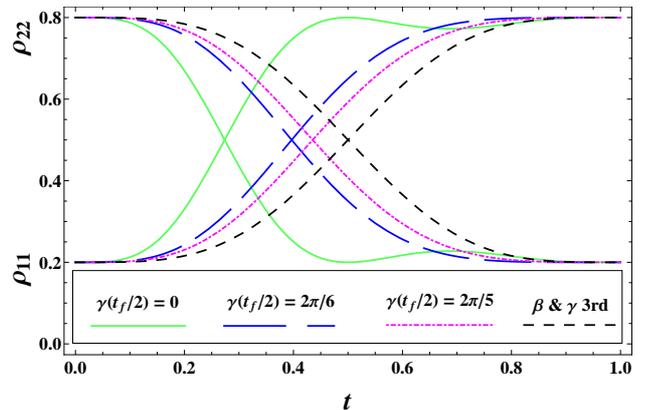}
\end{center}
 \caption{(Color online.) IEC control passages for different intermediate conditions in the legend on $\gamma$ at $t_f/2$.
 }
 \label{fig:usualUnusual}
\end{figure}

We did not have much freedom in manipulating $\gamma$ in the case where it was a third-order polynomial in $t$. It follows naturally to assume fourth-order $\gamma$, namely $\gamma(t)=\sum_{i=0}^4 b_i t^i$. To fix all the five coefficients in $\gamma$, one must add an extra condition to Eq.~(\ref{condGamma3rd}). We impose
conditions on the value of $\gamma(t_f/2)$ as
\begin{equation}
\text{(i)}\ \ \gamma(t_f/2)=2\pi/5,\ \ \ \text{(ii)}\ \ \gamma(t_f/2)=2\pi/6.
\label{condGamma4thMidNon0}
\end{equation}
Let us first consider the condition (i) in Eq.~(\ref{condGamma4thMidNon0}).
The fourth-order $\gamma$, with this new condition, again decreases monotonically from $\pi$ to $0$ as $t$ changes from $0$ to $t_f$. The regular behavior of the new $\gamma$ enables us to reuse the third-order $\beta$ obtained in the previous section. The corresponding IEC gives rise to control passages as shown by the dot-dashed curves in Fig.~\ref{fig:usualUnusual}, which are obviously faster than the dashed ones because the fidelity of the population inversion in the former approaches one faster. The latter ones are in fact the IEC passages in Fig.~\ref{fig:mixedresult3}. Therefore, this new fourth-order $\gamma$ expedites the IEC. 

\begin{figure}[h]
\begin{center}
\includegraphics[scale=0.58]{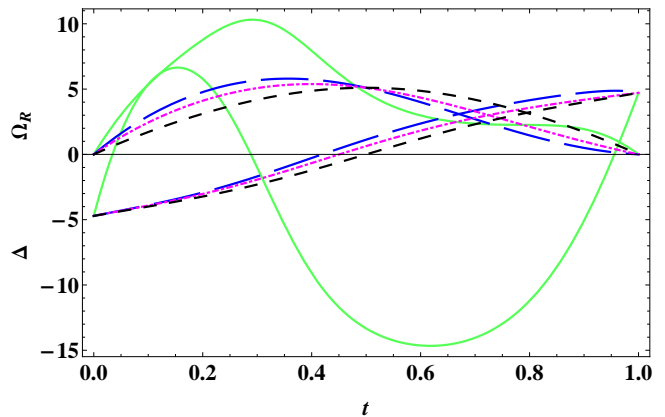}
\end{center}
 \caption{(Color online.) Parameters $\Omega_R$ (curves completely above the $t=0$ axis) and $\Delta$ (curves intersecting the $t=0$ axis) associated with the control passages in Fig.~\ref{fig:usualUnusual} respectively, with the same plot style.}
 \label{fig:usualUnusualOmegaDelta}
\end{figure}

In fact, further acceleration can be achieved by using the condition (ii) in 
Eq.~(\ref{condGamma4thMidNon0}) instead, which assigns $\gamma(t_f/2)$ a value 
smaller than that in the condition (i) in Eq.~(\ref{condGamma4thMidNon0}). 
This results in the control passages shown as the dashed curves in 
Fig.~\ref{fig:usualUnusual}. Figure \ref{fig:usualUnusualOmegaDelta} depicts the $\Omega_R$ and $\Delta$ corresponding to the solutions in Fig.~\ref{fig:usualUnusual}. One might wish to further speed up the control by tuning $\gamma(t_f/2)$ down continuously to some finite and positive value. Unfortunately, this is infeasible because the condition $\gamma(t_f/2)=2\pi/6$ is already near the limit, $2\pi/6.40175$, below which the fourth-order $\gamma$ that meets the condition behaves oddly in the sense that $\gamma(t)$ has a sharp drop into the negative domain at time near $t_f$. If $\beta$ remains third-order, singular behavior occurs in $\Delta$. The worse is that one cannot design a different $\beta$ that is able to diminish the singularity in $\Delta$ without introducing singular behavior in $\Omega_R$. 

\begin{figure}[h]
\begin{center}
\includegraphics[scale=0.45]{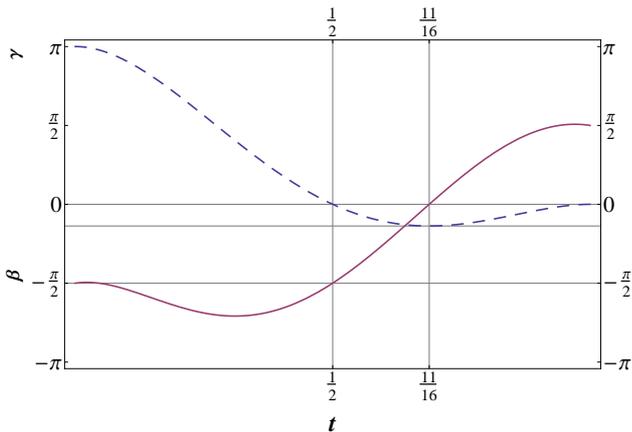}
\end{center}
\caption{(Color online.) Parameters $\gamma$ (dashed) and $\beta$ (solid) as solutions respectively to Eqs. (\ref{condGamma4thMid0}) and (\ref{condBeta5th4GammaMid0}). We set
$t_f=1$.}
\label{fig:usualUnusualBetaGamma}
\end{figure}

Nevertheless, there exist cases of IEC where certain singular behaviors in $\Omega_R$ and/or $\Delta$ caused by $\gamma$ can be remedied by $\beta$ if $\beta$ has some special profiles. When this happens, the IEC time significantly diminishes if another technique, to be
introduced shortly, is employed. We elucidate this by an example. Suppose we take $\gamma(t_f/2)=0$ and add this condition to Eq.~(\ref{condGamma3rd}) to 
impose a new set of conditions on $\gamma$ as
\begin{equation}\label{condGamma4thMid0}
\begin{aligned}
&\text{(i)}\ \gamma(0)=\pi,\ \ &\text{(ii)}\ \gamma(t_f)=0,\\
&\text{(iii)}\ \dot{\gamma}(0)=0,\ \ &\text{(iv)}\ \dot{\gamma}(t_f)=0,\\ &\text{(v)}\ \gamma(t_f/2)=0.
\end{aligned}
\end{equation}
Conditions in Eq.~(\ref{condGamma4thMid0}) yield a fourth-order solution of $\gamma$, as plotted in Fig.~\ref{fig:usualUnusualBetaGamma} (dashed curve).

As seen in Fig.~\ref{fig:usualUnusualBetaGamma}, $\gamma(t)$ reaches zero at $t_f/2$, and $\dot{\gamma}(t)$ changes sign at $11t_f/16$. The former makes $\cot\gamma$ blow up at $t_f/2$, while the latter may flip the sign of $\Omega_R$ 
by Eq.~(\ref{eqOmegaDelta}). Nonetheless, these issues can be made harmless by designing $\beta$ that satisfies the following conditions.
\begin{equation}\label{condBeta5th4GammaMid0}
\begin{aligned}
&\text{(i)}\ \beta(0)=-\pi/2,\ \ &\text{(ii)}\ \beta(t_f)=\pi/2,\\
&\text{(iii)}\ \beta(t_f/2)=-\pi/2,\ \ &\text{(iv)}\ \beta(11t_f/16)=0\\
&\text{(v)}\ \dot{\beta}(0)=-\dot{\beta}(t_f)=\pi/2t_f.
\end{aligned}
\end{equation}
Condition (iii) in Eq.~(\ref{condBeta5th4GammaMid0}) keeps the term $\cot\gamma\cos\beta$ in Eq.~(\ref{eqOmegaDelta}) finite at $t_f/2$ despite $\gamma(t_f)=0$. Notice from Eq.~(\ref{condGamma4thMid0}) that $\lim_{t\rightarrow 0}\dot{\gamma}=0^-$ and $\lim_{t\rightarrow t_f/2}\dot{\gamma}=0^+$. Conditions (i) and (ii) in Eq.~(\ref{condBeta5th4GammaMid0}) make $\dot{\gamma}/\sin\beta$, i.e., $\Omega_R$, approach zero from above in these cases. Because $\beta$ changes sign with $\dot{\gamma}$ at $11t_f/16$, $\Omega_R$ remains nonnegative over all time. Moreover, condition (iv) in Eq.~(\ref{condBeta5th4GammaMid0}) and that $\dot{\gamma}(11t_f/16)=0$ ensure $\Omega_R$ be finite at $11t_f/16$. Two more conditions should present themselves to fix the detuning at $t=0$ and $t=t_f$, which for this example we choose to be 
Eq.~(\ref{condBeta5th4GammaMid0})~(v). This particular choice seems to have no {\it a priori} reason over all possible choices. Indeed, as to be seen 
in Section \ref{secIEC}.5, the value of $\dot{\beta}(0)$ plays a key role in optimizing the energy cost of IEC, and Eq.~(\ref{condBeta5th4GammaMid0})~(v) is not the optimal choice. 

The solid curve in Fig.~\ref{fig:usualUnusualBetaGamma} depicts the solution of $\beta$ under the conditions Eq.~(\ref{condBeta5th4GammaMid0}), which demonstrates the explanations just given. The $\gamma$ and $\beta$ in Fig.~\ref{fig:usualUnusualBetaGamma} produce $\Omega_R$ and $\Delta$ as the solid curves in Fig.~\ref{fig:usualUnusualOmegaDelta} and the corresponding control passages (solid) in Fig.~\ref{fig:usualUnusual}. Interestingly, the solid control passages in Fig.~\ref{fig:usualUnusual} appears to attain population inversion with fidelity 1
at $t=t_f/2$, much earlier than all others in the figure, notwithstanding the fidelity drops after $t=t_f/2$ and re-approaches one at $t_f$. 

Does this surprising phenomenon truly give the population inversion as that to be achieved at $t_f$? The answer is ``Yes'', provided that anther technique is employed, which is switching the total Hamiltonian $H(t)$ at $t=t_f/2$ sharply to some constant one in which the population inversion is well defined. 
Population inversion of a two-level system is well-defined only by the Hamiltonian of the system without external fields. As such, the basis of the initial and final states of the control must be the eigenstates of $H_0$. Since $\sin[\beta(t_f)]$ is finite, conditions (iii) and (iv) in Eq.~(\ref{condGamma4thMid0}) sets $\Omega_R(0)$ and $\Omega_R(t_f)$ to zero, so that $[H(0),I(0)]=[H(t_f),I(t_f)]=0$. Accordingly, each eigenstate of $I(0)$ and that of $I(t_f)$ are also an eigenstate of $H(0)$ and one of $H(t_f)$, respectively. As we began the control with the state $\rho^I(0)=\mathrm{diag}((1-p_++p_-)/2,(1+p_+-p_-)/2))$, if we follow the solid control passage in Fig.~\ref{fig:usualUnusual} all the way to $t_f$, we end up with the state $\rho^I(t_f)=\mathrm{diag}((1+p_+-p_-)/2,(1-p_++p_-)/2))$, which is the very population-inverted mixed state in the two-level basis of $H_0$. 

If, on the other hand, the evolution halts at $t=t_f/2$ along the solid control passage in Fig.~\ref{fig:usualUnusual}, the mixed state in the eigenbasis of
$I(t_f/2)$ is $\rho^I(t_f/2)=\mathrm{diag}((1+p_+-p_-)/2,(1-p_++p_-)/2))$, the same as $\rho^I(t_f)$. Nonetheless, we only impose $\gamma(t_f/2)=0$ in Eq.~(\ref{condGamma4thMid0}) but not $\dot{\gamma}(t_f/2)=0$. Thus, $[H(t_f/2),I(t_f/2)]\neq 0$, and $\rho^I(t_f/2)$ is not the right population-inverted state in the eigenbasis of the Hamiltonian.

Fortunately, this issue has a resolution. In many applications of quantum control, such as in NMR, the total Hamiltonian $H(t)$ is actually piecewise constant or piecewise continuous. Therefore, at $t_f/2$, one can stop the control passage of the system and instantaneously switch the total Hamiltonian to a constant diagonal Hamiltonian in order that it commutes with $I(t_f/2)$, which is also diagonal. To do so, we simply need to turn off the RF field right at $t=t_f/2$, i.e., setting $\Omega_R(t > t_f/2)=0$, then the Hamiltonian becomes $H(t_f/2)=\mathrm{diag}(\Delta(t_f/2),\ -\Delta(t_f/2))$, which shares the basis with $\rho^I(t_f/2)$; hence, a well-defined population inversion completes. Note that $\Delta(t_f/2)\neq\Delta(t_f)$ in general but this does not affect the definition of population inversion unless $\Delta(t_f/2)=0$. Therefore, the population inversion at $t_f/2$ along the solid control passage in Fig.~\ref{fig:usualUnusual} should be authentic, which exhibits a tremendous decrement in the control time. We christen the control passages such as the solid curves in Fig.~\ref{fig:usualUnusual}, owing to conditions as in Eq.~(\ref{condGamma4thMid0}) and alike, the {\bf antedated control passages} to distinguish them from the usual ones that end at $t_f$. 

One may wonder why the antedated control passages are meaningful, as one can simply set the control time of the usual control passages to be some $t_f'<t_f$, say, $t_f'=t_f/2$, given the aforementioned fact that $t_f$ can be anything mathematically. Here is our reasoning. On the one hand, for whatever $t_f$ chosen for the usual control passages, the antedated ones can always complete the control much earlier. On the other hand, the energy cost due to the antedated control passages can be made smaller than that due to the usual ones. Actually, all the usual IEC passages in Fig.~\ref{fig:usualUnusual} have the same energy cost, namely $\int_0^{t_f'}\Omega_R(t)\mathrm{d}t\approxeq 3.1482$. One may keep this number in mind to compare it with the energy cost of the antedated control passages
introduced in the next section.
 
The antedated control passages speed up the control at a price of weakened stability. As seen in Fig.~\ref{fig:usualUnusual}, the population inversion achieved by the antedated control passages (solid) appears at the peak and valley of the passages; hence, as opposed to the case of usual control passages, the evolution of the state must halt sharply at $t_f/2$ because somewhat off that specific temporal position would reduce the fidelity. This price certainly depends on how much tolerance in compromising the fidelity is practical.

We remark that although switching between Hamiltonians takes some amount of 
time, turning on or off a RF field in NMR costs negligible time compared with the control time of IEC \cite{nmr}. 

In the next subsection, we will look for more antedated control passages that are even faster and study their optimization in terms of control time and energy cost. 
\subsection{Optimizing IEC}
Condition (v) in Eq.~(\ref{condGamma4thMid0}) urges one to modify it by merely shifting the antedated time $t_a$, at which $\gamma$ hits zero, to attain faster antedated control passages.  The time $t_a$ can be even earlier than $t_f/2$. 
Instead of the condition (v) in Eq.~(\ref{condGamma4thMid0}), we consider
three different intermediate conditions:
\begin{equation}\label{condGamma4thInt0}
\mathrm{(v')}\ \gamma(t_f/3)=0, \ \ \mathrm{(v'')}\ \gamma(2t_f/7)=0,\ \ \mathrm{(v''')}\ \gamma(2t_f/7.6)=0. 
\end{equation}
\begin{figure}[h]
\begin{center}
\includegraphics[scale=0.41]{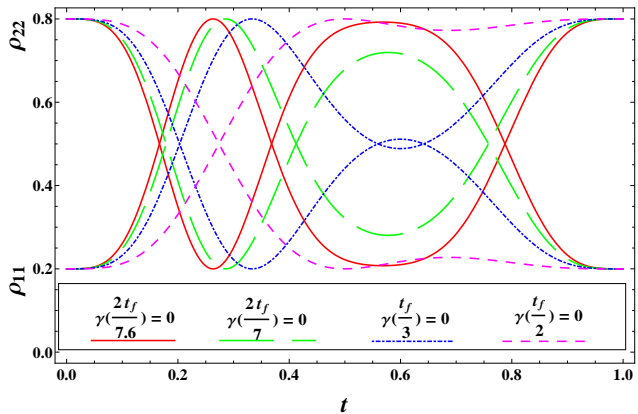}
\end{center}
\caption{(Color online.) Antedated control passages due to fourth-order $\gamma$ and fifth-order $\beta$. The antedated time of each passage is shown in the legend.}
\label{fig:population5thBeta}
\end{figure}

We take $\beta$ fifth-order in $t$ and require it to satisfy conditions adapted to each case in Eq.~(\ref{condGamma4thInt0}) separately, similarly to what we did to obtain Fig.~\ref{fig:usualUnusualBetaGamma}. Basically, conditions (iii) and (iv) in Eq.~(\ref{condBeta5th4GammaMid0}) should be modified in accordance with the three cases in Eq.~(\ref{condGamma4thInt0}). Nonetheless we choose not to show 
the new conditions on $\beta$ here, as they are irrelevant for our purpose at the moment. The antedated control passages obtained are shown in Fig.~\ref{fig:population5thBeta}, which demonstrate that the control time shortens but the stability of fidelity weakens with the decrease of the antedated time $t_a$. 

\begin{figure}[h!]
\begin{center}
\includegraphics[scale=0.4]{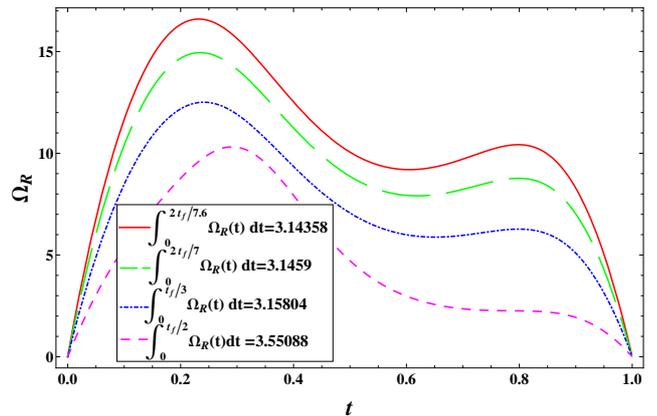}
\end{center}
\caption{(Color Online.) $\Omega_R$ and the energy cost correspond to each control passage in Fig.~\ref{fig:population5thBeta}.}
\label{fig:omega5thBeta}
\end{figure}

\begin{figure}[h!]
\begin{center}
\includegraphics[scale=0.36]{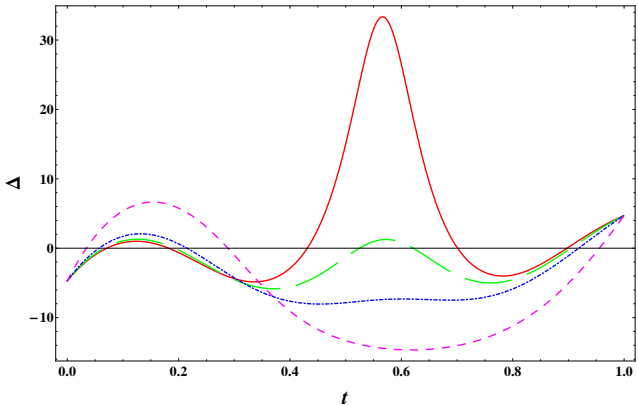}
\end{center}
\caption{(Color online.) Detuning corresponding to the IEC passages in Fig.~\ref{fig:population5thBeta}.}
\label{fig:delta5thBeta}
\end{figure}

Now let us concentrate on the profiles of $\Omega_R$ and $\Delta$ because they determine the cost of the control and physical realizability of the parameters $\gamma$ and $\beta$. Figure \ref{fig:omega5thBeta} illustrates $\Omega_R$ due to the condition (v) in Eq.~(\ref{condGamma4thMid0}) and the conditions in Eq.~(\ref{condGamma4thInt0}), while Fig.~\ref{fig:delta5thBeta} delineates the corresponding detunings. The parameters $\Omega_R$ and $\Delta$ in Figs.~\ref{fig:omega5thBeta} and \ref{fig:delta5thBeta} are obviously physical. 

To obtain the energy cost of each antedated control passage in Fig.~\ref{fig:population5thBeta}, we need to integrate each $\Omega_R$ in Fig.~\ref{fig:omega5thBeta} over time from $t=0$ up to the antedated time $t_a<t_f$, when the population inversion is completed. The legend in Fig.~\ref{fig:omega5thBeta} presents the result, which shows that the faster the antedated control passage the less the energy cost. This conclusion may be somewhat rush because we do not know yet if the energy cost of each antedated control passages under investigation is optimized. It turns out to be the case as we will show below.

The integrals shown in the legend of Fig.~\ref{fig:omega5thBeta} suggests the existence of a certain lower bound in the energy cost. Although it turns out to be true, merely decreasing the antedated time $t_a$ can hardly lead to the bound,
contrary to a naive expectation. We point out the reason without demonstration. By our study, as soon as $t_a$ gets less than $\sim 2t_f/7.7$, the corresponding $\gamma$ will not only cross zero at $t_a$ but also be below $-\pi$ at a later time, which brings to $\Omega_R$ and $\Delta$ extra unphysical behaviors, such as discontinuity in $\Omega_R$ and $\Delta$ and divergence in $\Delta$, which cannot be simultaneously remedied by any form of $\beta$. Nevertheless, for any fixed $t_a$, the energy cost admits a lower bound that is dictated by the value of $\dot{\beta}(0)$, as pointed out earlier.
\begin{figure}[h]
\begin{center}
\includegraphics[scale=0.49]{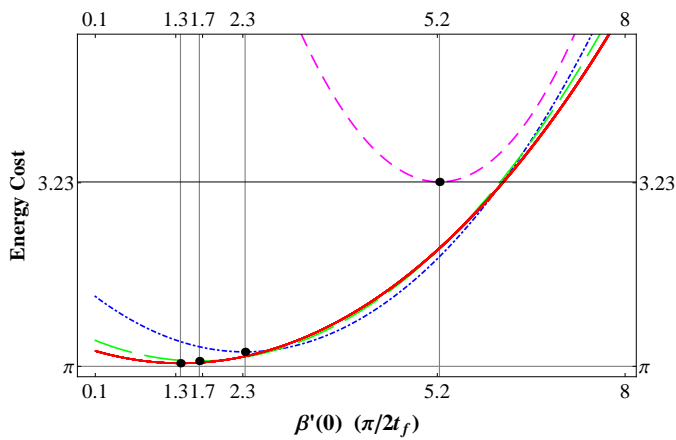}
\end{center}
\caption{(Color Online.) Energy cost as a function of $\dot{\beta}(0)$ (in units of $\pi/2t_f$). Each curve corresponds to the control passage in the same color in Fig.~\ref{fig:population5thBeta} with the specified $t_a$. Black dots on the curves mark the minima of the curves, whose values are recorded in Eq.~(\ref{eqECost}).}
\label{fig:energycost}
\end{figure}

To illuminate the relation between $\dot{\beta}(0)$ and the energy cost of an antedated control passage, we again consider the control passages in Fig.~\ref{fig:population5thBeta}, for each of which we sweep $\dot{\beta}(0)$ in a wide range, say $\dot{\beta}(0)\in[0.1\pi/2t_f,8\pi/2t_f]$, then study the energy cost of the control passage as a function of $\dot{\beta}(0)$. The result is plotted in Fig.~\ref{fig:energycost}. Seen in this figure, each curve of energy cost has a minimum at certain value of $\dot{\beta}(0)$, as marked by the black dots therein. Note that $\dot{\beta}(t_f)$ is different from $\dot{\beta}(0)$ by only a minus sign, see e.g., Eq.~(\ref{condBeta5th4GammaMid0})~(v). We collect the minimal energy costs in Fig.~\ref{fig:energycost} as follows.
\begin{equation}\label{eqECost}
\begin{aligned}
& \int_0^{t_a={2t_f}/{7.6}}\Omega_R\mathrm{d}t\Bigr|_{\min}=3.143,\ \ \dot{\beta}(0)=\frac{1.37\pi}{2t_f},\\
& \int_0^{t_a={2t_f}/{7}}\Omega_R\mathrm{d}t\Bigr|_{\min}=3.144,\ \ \dot{\beta}(0)=\frac{1.652\pi}{2t_f},\\
& \int_0^{t_a={t_f}/{3}}\Omega_R\mathrm{d}t\Bigr|_{\min}=3.149,\ \ \dot{\beta}(0)=\frac{2.334\pi}{2t_f},\\
& \int_0^{t_a={t_f}/{2}}\Omega_R\mathrm{d}t\Bigr|_{\min}=3.230,\ \ \dot{\beta}(0)=\frac{5.232\pi}{2t_f}.
\end{aligned}
\end{equation}

According to Fig.~\ref{fig:energycost}, the order of the energy costs in the legend of Fig.~\ref{fig:omega5thBeta} is no more than a coincidence that arises for our unintended choice of $\dot{\beta}(0)=\pi/2t_f$ for those cases. Despite this, Eq.~(\ref{eqECost}), which is extracted from Fig.~\ref{fig:energycost}, verifies that the energy cost of an antedated control passage reduces with the antedated time $t_a$, as long as the right $\dot{\beta}(0)$ is imposed to minimize the energy cost of the control passage for a given value of $t_a$. Therefore, antedated control passages can be optimized by manipulating both $t_a$ and $\dot{\beta}(0)$. 

Equation~(\ref{eqECost}) also points to the existence of a lower bound
$\pi$ of the minimal energy cost of antedated IEC. In other words, an IEC is extremely optimized in terms of energy cost when the inversely-engineered Rabi frequency $\Omega_R$ happens to be a $\pi$-pulse, as $\int_0^{t_a}\Omega_R\mathrm{d}t=\pi$. Nevertheless, one must diminish $t_a$ further, beyond $2t_f/7.6$, to achieve this extremal case. However, this will result in unphysical $\Delta$ and $\Omega_R$ before the extreme is hit as explained before. 

Now that the energy cost of the antedated control passages of IEC can be optimized while the detuning frequency is physical, stability of the fidelity appears to be the only constraint apart from the uncertainty principle and the Hamiltonian-switching time. The last condition depends on how sharply one can terminate the evolution of the system. 
Given the available experimental power nowadays, there would not be any problem 
in switching the Hamiltonian in the scale of nanoseconds in NMR, for example. 
\section{Conclusions and Outlook}
In conclusion, we have shown by taking the two-level system as an example, that
IEC indeed offers fast control passages --- in particular the antedated control passages --- as shortcuts to adiabatic control of quantum systems. To design a LRI for a system, appropriate boundary conditions on the parameters in the LRI can be imposed to meet various goals of the control. In the scheme that is based on our polynomial ansatz and the family of boundary conditions of $\beta$ and $\gamma$, an IEC passage is optimal in terms of speed and energy cost when the inversely-engineered Rabi frequency obtained from the designed LRI is approximately a $\pi$-pulse. We expect to study the optimization of IEC more systematically by means of control theory. All our results apply to pure states, which are merely a special case of our general analysis. Besides, we note that Carlini {\it et al}. \cite{CarliniPRL2006} studied time-optimal quantum evolution and the optimal Hamiltonian for a quantum system via variational principle. It would be interesting to see if and how their results are related to the optimization of IEC.

Mixed states are the subject of IEC in this paper, it would be intriguing to apply IEC to NMR quantum computation, in which mixed states are used for computation, to further shorten the operation time of the one-qubit quantum gates implemented by RF-pulses in NMR.

It is also natural to extend IEC to multi-level systems. To this end, one has to first design reasonable and most convenient LRIs for these systems, whose forms are rather hard to guess for complicated systems, which is under our investigation and will be reported elsewhere.
\section*{Acknowledgements}
We appreciate Xi Chen for discussion. We also owe gratitudes to Utkan G\"ung\"ord\"u for discussion and his comments on the drafts of the paper. We thank the support from ``Open Research Center'' Project for Private Universities: matching fund subsidy from MEXT, Japan. 
MN would like to thank partial supports of Grants-in-Aid for Scientific 
Research from the JSPS (Grant No.~23540470).

\end{document}